\documentclass[12pt]{article}
\usepackage{amsmath,amssymb,amsfonts}
\usepackage{geometry}
\usepackage{color}
\usepackage{doi}
\title{\textbf{An Extended 2+1-Dimensional Gardner Equation: On Moving Boundary Problems Solvable via Ermakov-Painlev\'e II Symmetry Reduction}}
\author{
    Colin Rogers$^1$ and Sandra Carillo$^{2, 3}$ \footnote{\bf corresponding author,   email: sandra.carillo@uniroma1.it} \\
    \small $^1$University of New South Wales, Sydney, New South Wales, Australia \\
    \small $^2$ Dipartimento Scienze di Base e Applicate per l'Ingegneria \\
    \small Universit\`a di Roma  ``La Sapienza'', 16, Via A. Scarpa, 00161 Rome, Italy \\
    \small $^3$ Gr. Roma1, IV - Mathematical Methods in NonLinear Physics  (MMNLP)\\
    \small National Institute for Nuclear Physics (I.N.F.N.), Rome, Italy
}
\date{\today}

\begin{document}

\maketitle

\begin{abstract}
A   class of moving boundary problems of Stefan-type for an extension of the 2+1- dimensional Gardner equation is shown to be amenable to exact solution by application of a Ermakov- Painlev\'e II symmetry reduction.
\end{abstract}

\section*{I Introduction}

Interconnections between canonical 2+1-dimensional solitonic equations and their hierarchies, notably via reciprocal transformations, have been detailed in \cite{1,2,{2b}}. In \cite{3}, a 2+1-dimensional extended Dym equation has recently been introduced together with its Lax pair, $\bar{\partial}$-bar dressing scheme and temporal modulation by means of involutory transformations with genesis in the theory of Ermakov systems. In \cite{4}, classes of nonlinear moving boundary problems of Stefan-type for the 1+1-dimensional solitonic Gardner equation and a novel reciprocal associate have been shown to be amenable to exact solution via Painlev\'e II symmetry reduction. Here, hybrid Ermakov-Painlev\'e II symmetry reduction is established for a novel class of 2+1-dimensional extended Gardner equations and moving boundary problems for the latter solved thereby.

Ermakov-Painlev\'e II systems had their origin in \cite{5} in connection with wave packet representations admitted by multi-dimensional coupled nonlinear Schr\"odinger systems which incorporate a de Broglie-Bohm potential term. The canonical single component Ermakov-Painlev\'e II equation was derived therein, in particular in the context of transverse wave propagation in a generalised Mooney-Rivlin hyperelastic material. It has subsequently arisen in such diverse physical applications as cold plasma physics \cite{6}, Korteweg capillary theory \cite{7} and in the analysis of Dirichlet boundary value problems associated with the classical Nernst-Planck electrodiffusion system \cite{8}. Underlying integrable structure in multi-component Ermakov-Painlev\'e II systems has been delimited in \cite{9} and an admitted B\"acklund transformation applied iteratively therein.

\section*{II A novel extended 2+1-dimensional Gardner equation}

The 2+1-dimensional Gardner equation as originally set down in \cite{10}, (see \cite{11}) is
\begin{equation}
v_t + v_{xxx} + 6 \beta_0 v v_x - 3 \beta_1^2 v^2 v_x + 3 \partial_x^{-1} v_{yy} - 3 \beta_1 v_x \partial_x^{-1} v_y = 0
\tag{2.1}
\end{equation}
has subsequently been the subject of extensive investigation. Here, an extended version of this 2+1-dimensional Gardner equation is under consideration, namely
\begin{equation}
v_t + v_{zzz} + 6 \beta_0 v v_z -( 3/2) \beta_1^2 v^2 v_z + 3 \partial_z^{-1} v_{yy} - 3 \beta_1 v_z \partial_z^{-1} v_y 
+ \lambda ( t + a)^{-2} (v + \delta)^{-4} v_z = 0 .
\tag{2.2}
\end{equation}

In \cite{12}, a reduction of the standard 1+1-dimensional Gardner equation with $\beta_0 = 1$, $ \beta_1^2 = 4$ to the canonical solitonic mKdV equation was obtained on introduction of
\begin{equation}
\begin{array}{ll}
 v = 1/2 + u,\\
x = -\frac{3}{2} \tau+ z, \quad t = \tau
\end{array}
\tag{2.3}
\end{equation}
so that
\begin{equation}
\left.\partial_\tau\right|_z =- \frac{3}{2} \partial_x + \partial_t, \quad \left.\partial_z\right|_\tau = \partial_x .
\tag{2.4}
\end{equation}

A novel extension of the mKdV equation which incorporates an additional temporally modulated term of the type in (2.2) has recently been shown in \cite{13} to admit Ermakov-Painlev\'e II symmetry reduction. Thereby, in particular, a class of moving boundary problems of Stefan-type has been solved explicitly.

In the present 2+1-dimensional context, application of (2.3) to (2.2) results with $\delta = -1/2$ in
\begin{equation}
\begin{aligned}
u_t - 3/2 u_x + u_{xxx} + 6 \beta_0 (1/2 + u) u_x - (3/2)  \beta_1^2 (1/2 + u)^2 u_x \nonumber \\
+ 3 \partial_z^{-1} u_{yy} - 3 \beta_1 u_x \partial_z^{-1} u_y + \lambda ( t + a)^{-2} u^{-4} u_x = 0
\end{aligned}
\tag{2.5}
\end{equation}
where (2.4)$_2$ shows that $\partial_z^{-1} u_{yy} = \partial_x^{-1} u_{yy}$ and $\partial_z^{-1} u_y = \partial_x^{-1} u_y$. Accordingly, (2.5) becomes
\begin{equation}
\begin{aligned}
u_t + [-3/2 + 3 \beta_0 - 3/8 \beta_1^2] u_x + u_{xxx} + [6 \beta_0 - 3/2 \beta_1^2] u u_x - 3/2 \beta_1^2 u^2 u_x \nonumber \\
+ 3 \partial_x^{-1} u_{yy} - 3 \beta_1 u_x \partial_x^{-1} u_y + \lambda ( t + a)^{-2} u^{-4} u_x = 0~.
\end{aligned}
\tag{2.6}
\end{equation}

\section*{III Ermakov-Painlev\'e II symmetry reduction}

A symmetry reduction ansatz is now introduced according to
\begin{equation}
u = ( t + a)^m \Psi\left(\frac{x + \eta y}{( t + a)^n}\right), \quad \eta \in \mathbb{R}
\tag{3.1}
\end{equation}
with implied relations
\begin{equation}
\partial_x^{-1} u_y = \eta u, \quad \partial_x^{-1} u_{yy} = \eta^2 u_x
\tag{3.2 , 3.3}
\end{equation}
whence (2.6) yields
\begin{equation}
\begin{aligned}
u_t + [-3/2 + 3 \beta_0 - 3/8 \beta_1^2 + 3 \eta^2] u_x + u_{xxx} + [6 \beta_0 - 3/2 \beta_1^2 - 3 \beta_1 \eta]~ u u_x \nonumber \\
- 3/2 \beta_1^2 u^2 u_x + \lambda ( t + a)^{-2} u^{-4} u_x = 0~.
\end{aligned}
\tag{3.4}
\end{equation}
On insertion of (3.1) into the latter, there results
\begin{equation}
\begin{aligned}  
m   ( t + a)^{m-1} \Psi + ( t + a)^{m-1} (-n \xi) \Psi' + ( t + a)^{m-n} [-3/2 + 3 \beta_0 - 3/8 \beta_1^2 + 3 \eta^2] \Psi' \nonumber \\
+ ( t + a)^{m-3n} [ \Psi''' - 3/2 \beta_1^2 \Psi^2 \Psi']+ ( t + a)^{2m-n} [6 \beta_0 - 3/2 \beta_1^2 - 3 
 \beta_1 \eta] \Psi \Psi' 
  \nonumber \\
+ \lambda ( t + a)^{-2-3m-n} \Psi^{-4} \Psi' = 0
\end{aligned}
\tag{3.5}
\end{equation}

wherein
$$\xi = (x + \eta y) / ( t + a)^n .$$

Thus, with $m = -1/3, n = 1/3$ and $\beta_0, \beta_1$ determined by imposition of the pair of  {{imposed}}
conditions
\begin{equation}
\eta = \frac{2 [ \beta_0 - \beta_1^2 /4] }{ \beta_1}, \quad \eta^2 = 1/2 - \beta_0 + \beta_1^2 / 8
\tag{3.6}
\end{equation}
(3.5) reduces to
\begin{equation}
\Psi^{\prime\prime\prime} - (3/2) \beta_1^2 \Psi^2 \Psi^\prime - 1/3  ( \xi \Psi)^\prime + \lambda \Psi^{-4} \Psi^\prime = 0 .
\tag{3.7}
\end{equation}
On integration of the latter
\begin{equation}
\Psi'' - (1/2) \beta_1^2 \Psi^3 - (1/3) \xi \Psi - ( \lambda/3)  \Psi^{-3} = \mathbb{K}, \quad \mathbb{K} \in \mathbb{R} ,
\tag{3.8}
\end{equation}
which, on appropriate scaling: $\Psi = \mu w^*, \xi = \nu \xi^*$ and with $\mathbb{K} = 0$ results in the canonical base Ermakov-Painlev\'e II equation as introduced in \cite{5}, namely
\begin{equation}
w^*_{\xi^* \xi^*} = 2 w^{*3} + \xi^* w^* + \delta^* w^{*-1}, \quad \delta^* \in \mathbb{R} .
\tag{3.9}
\end{equation}
A connection between the above Ermakov-Painlev\'e II equation and the classical Painlev\'e XXXIV equation has been detailed in \cite{9}.

Hybrid Ermakov-Ray-Reid / Painlev\'e II systems introduced therein have recently been applied in the symmetry reduction and exact solution thereby of moving boundary problems for a novel mKdV-type coupled system.

\section*{IV A class of moving boundary problems}

Imposition in (3.4) of the constraints in (3.6) so that
\begin{equation}
1/2 - \beta_0 + 1/8 \beta_1^2 = 4 [\beta_0 - \beta_1^2 / 4]^2 / \beta_1^2
\tag{4.1}
\end{equation}
results in an extended mKdV-type equation
\begin{equation}
u_t + u_{xxx} - (3/2) \beta_1^2 u^2 u_x + \lambda ( t + a)^{-2} u^{-4} u_x = 0~.
\tag{4.2}
\end{equation}
Therein, the variable $y$ enters implicitly via the relation (3.1).

A class of moving boundary problems for the latter is now considered with the conditions
\begin{equation}
\left. \begin{array}{l} u_{xx} - (\beta_1^2 / 2)  u^* - (\lambda/3) ( t + a)^{-2} u^{-3} = L_m S^i\dot S \\ \\ u = P_m S^i \end{array} \right\} \text{ on } x + \eta y = S(t), \quad t > 0
\tag{4.3}
\end{equation}
$$\left. [ u_{xx} - (\beta_1^2 / 2)  u^3 - (\lambda/3) ( t + a)^{-2} ] \right|_{x+\eta y=0} = H_0 ( t + a)^{\kappa}$$
wherein
$$S(t) = \gamma ( t + a)^{1/3} .$$

\subsubsection*{Boundary Conditions}

\paragraph{I}
$$u_{xx} - (\beta_1^2 / 2)   u^3 - (\lambda/3) ( t + a)^{-2} u^{-3} = L_m S^i \dot S \quad \text{on } x + \eta y = S(t), \quad t > 0$$
Insertion of the 2+1-dimensional symmetry ansatz (3.1) with $m = -1/3, n = 1/3$ into the preceding yields
\begin{equation}
\Psi''(\gamma) - (\beta_1^2 / 2)  \Psi^3(\gamma) 
- (\lambda/3) \Psi^{-3}(\gamma) = (1/3) L_m
\tag{4.4}
\end{equation}
together with $i = -1$. By virtue of (3.8) with $\mathbb{K} = 0$
\begin{equation}
\Psi''(\gamma) - (\beta_1^2 / 2) \lambda \Psi^3(\gamma) - (1/3) \gamma \Psi(\gamma) - (\lambda/3) \Psi^{-3}(\gamma) = 0
\tag{4.5}
\end{equation}
whence (4.4) yields
\begin{equation}
L_m = \gamma \Psi(\gamma)~ .
\tag{4.6}
\end{equation}

\paragraph{II}
$$u =P_m S^i \quad \text{on } x + \eta y = S(t), \quad t > 0$$
The above implies $i = -1$ together with
\begin{equation}
P_m = \gamma \Psi(\gamma)~ .
\tag{4.7}
\end{equation}

\paragraph{III}
$$\left. [u_{xx} - (\beta_1^2 / 2) u^3 - (\lambda/3) ( t + a)^{-2} u^{-3} ] \right|_{x+\eta y=0} = H_0 ( t + a)^{\kappa}, \quad t > 0~.$$
This yields
$$\Psi''(0) - (\beta_1^2 / 2) \Psi^3(0) - (\lambda/3) \Psi^{-3}(0) = H_0 ( t + a)^{\kappa + 1}$$
so that $\kappa = -1$ and 
%
on use of (3.8) with $\mathbb{K} = 0$ at $\xi = 0$.

It is remarked that exact solution $\Psi$ of the Ermakov-Painlev\'e II equation which possesses Airy-type representation have been applied in a Korteweg capillary system context in \cite{7}.

\section*{V Temporal modulation}
Involutory-type transformations with genesis in connection with coupled Ermakov-Ray-Reid systems in \cite{15} have subsequently been applied {\it mutatis mutandis} in modern soliton theory to the systematic reduction of certain spatially modulated coupled sine-Gordon,  Demoulin and Manakov-type   to their unmodulated S-integrable counterparts \cite{16}. The modulated systems thereby inherit key properties of their canonical unmodulated associates such as admittance of invariance under B\"acklund transformations \cite{17, 18} and being amenable to the inverse scattering procedure \cite{19}.

Here, the class of transformations
\begin{equation}
 \begin{aligned}
 d t^* = \rho^{-2}(t) d t, \quad d x^* = d x, \quad d y^* = d y,   \cr
\rho^* = 1/\rho(t), \quad v^* = v/\rho(t)
\end{aligned}  \quad { I}^*
\tag{5.1}
\end{equation}
\noindent
which admits the involutory property $I^{**} = I$ is applied to the extended 2+1-dimensional Gardner equation
\begin{equation}
\begin{aligned}
v_t + v_{xxx} + 6 \beta_0 v v_x - 3/2 \beta_1^2 v^2 v_x + 3 \partial_x^{-1} v_{yy} - 3 \beta_1 v_x \partial_x^{-1} v_y \nonumber \\
+ \lambda ( t + a)^{-2} (v + \delta)^{-4} v_x = 0
\end{aligned}
\tag{5.2}
\end{equation}
There results an associated diverse class with temporal modulation, namely
\begin{equation}
\begin{aligned}
\rho^{*2} \partial_{t^*} (\rho^{*-1} v^{*}) + \rho^{*-1} v^*_{x^*x^*x^*} + 6 \beta_0 \rho^{*-2} v^* v^*_{x^*} - 3/2 \beta_1^2 \rho^{*-3} v^{*2} v^*_{x^*} \nonumber \\
+ 3 \rho^{*-1} \partial_{x^*}^{-1} v^*_{y^*y^*} - 3 \beta_1 \rho^{*-2} v^*_{x^*} \partial_{x^*}^{-1} v^*_{y^*} \nonumber \\
+ \lambda (\partial_{t^*}^{-1} \rho^{*-2} + a)^{-2} (v^* \rho^{*-1} + \delta)^{-4} \rho^{*-1} v^*_{x^*} = 0 .
\end{aligned}
\tag{5.3}
\end{equation}

In \cite{3}, the involutory-type transformations have recently been applied to a novel 2+1-dimensional extension of the classical solitonic Dym equation. Therein, $\rho^*$ in (5.1) was determined by the Ermakov equation with its admitted nonlinear superposition principle (see \cite{20} and literature cited therein). Application of the latter may likewise be applied in (5.3) to generate a wide class of modulated versions of the 2+1-dimensional extended Gardner equation (5.2) which inherit properties of their unmodulated counterparts by virtue of the involutory property $I^{**}  = I$.
\subsection*{Acknowledgements}

S.C.\ acknowledges:
\begin{itemize}
\item Gr. Roma1, IV - Mathematical Methods in NonLinear Physics (MMNLP), National Institute for Nuclear Physics (I.N.F.N.), Rome, Italy;
\item Dipartimento di Scienze di Base e Applicate per l'Ingegneria,
Università di Roma  ``La Sapienza'', Rome, Italy;
\item Gruppo Nazionale di Fisica Matematica (G.N.F.M- I.N.D.A.M.), Italy;
\item Regione Lazio,   project CTE, Italy
\end{itemize}
 for support of the present research activity.


\begin{thebibliography}{99}

\bibitem{1} W. Oevel,  C. Rogers, Gauge transformations and reciprocal links in 2+1-dimensions, \textit{Rev. Math. Phys.} \textbf{5}, 299--330 (1993).

\bibitem{2} C. Rogers, The Harry Dym equation in 2+1-dimensions: a reciprocal link with the Kadomtsev--Petviashvili equation, \textit{Phys. Lett. A} \textbf{120}, 15--18 (1987).

\bibitem{2b}{W. Oevel, S.  Carillo, \ {\sl Squared
Eigenfunction Symmetries for Soliton  Equations:
Part I}, \ \textit{J.  Math. Anal. and Appl.}, {\bf 217},
 161--178,  (1998). ~~\doi{10.1006/jmaa.1997.5707}}
\bibitem{3} B.G. Konopelchenko, C. Rogers,  P. Amster, On an integrable 2+1-dimensional extended Dym equation: Lax pair, $\bar{\partial}$-dressing scheme and modulation, \textit{Open Communications in Nonlinear Mathematical Physics} \textbf{5}, 1--14 (2026).

\bibitem{4} C. Rogers, On moving boundary problems for the solitonic Gardner equation: a reciprocally associated class, \textit{Zeit. angew. Math. Phys.} \textbf{26}, 180 (2025).

\bibitem{5} C. Rogers, A novel Ermakov--Painlev\'e II system: 1+1-dimensional coupled NLS and elastodynamic reductions, \textit{Stud. Appl. Math.} \textbf{133}, 214--231 (2014).

\bibitem{6} C. Rogers,  P.A. Clarkson, Ermakov--Painlev\'e II reduction in cold plasma physics: Application of a B\"acklund transformation, \textit{J. Nonlinear Mathematical Physics} \textbf{25}, 247--261 (2018).

\bibitem{7} C. Rogers,  P.A. Clarkson, Ermakov--Painlev\'e II symmetry reduction in a Korteweg capillary system, \textit{Symmetry, Integrability and Geometry: Methods and Applications} \textbf{13}, 018 (2017).

\bibitem{8} P. Amster,  C. Rogers, On a Ermakov--Painlev\'e II reduction in three-ion electrodiffusion: A Dirichlet boundary value problem, \textit{Discrete and Continuous Dynamical Systems} \textbf{35}, 3277--3292 (2015).

\bibitem{9} C. Rogers,  W.K. Schief, On Ermakov--Painlev\'e II systems: Integrable reduction, \textit{Meccanica} \textbf{51}, 2957--2974 (2016).

\bibitem{10} B.G. Konopelchenko,  V.G. Dubrovsky, B\"acklund--Calogero group and general form of integrable equations for the two-dimensional Gelfand--Dikij--Zakharov--Shabat spectral problem: bilocal approach, \textit{Physica D} \textbf{16}, 79 (1985).

\bibitem{11} B.G. Konopelchenko, \textit{Introduction to Multidimensional Integrable Equations. The Inverse Spectral Transform in 2+1 Dimensions}, Technical Editor C. Rogers, Plenum Press, New York and London (1992).

\bibitem{12} A.V. Slyunyaev,  E.N. Pelinovsky, Dynamics of large amplitude solitons, \textit{J. Experimental \& Theoretical Physics} \textbf{89}, 173--181 (1999).

\bibitem{13} C. Rogers,  A.C. Briozzo, Moving boundary problems for a novel extended mKdV equation. Application of Ermakov--Painlev\'e II symmetry reduction, \textit{Open Communications in Nonlinear Mathematical Physics} \textbf{5}, 116--130 (2025).

\bibitem{14} C. Rogers,  A.C. Briozzo, Hybrid Ermakov--Ray--Reid / Painlev\'e II symmetry reduction: application to a class of moving boundary problems, \textit{Meccanica} \textbf{61}, 110 (2026).

\bibitem{15} C.Athorn, C. Rogers, U. Ramgulam,  A. Osbaldestin, On linearization of the Ermakov system, \textit{Phys. Lett. A} \textbf{143}, 207--212 (1990).

\bibitem{16} C. Rogers, W.K. Schief,  B. Malomed, On modulated coupled systems: Canonical reduction via reciprocal transformation
s, \textit{Communications Nonlinear Science \& Numerical Simulation} \textbf{83}, 105091 (2020).

\bibitem{17} C. Rogers,  W.F. Shadwick, \textit{B\"acklund Transformations and Their Applications}, Academic Press, Mathematics in Science and Engineering Series, New York (1982).

\bibitem{18} C. Rogers,  W.K. Schief, \textit{B\"acklund and Darboux Transformations: Geometry and Modern Applications in Soliton Theory}, Cambridge Texts in Applied Mathematics, Cambridge University Press (2002).

\bibitem{19} M.J. Ablowitz,  P.A. Clarkson, \textit{Solitons, Nonlinear Evolution Equations and Inverse Scattering}, Cambridge University Press (1991).

\bibitem{20} C. Rogers,  U. Ramgulam, A nonlinear superposition principle and Lie group invariance: application in rotating shallow water theory, \textit{Int. J. Non-Linear Mechanics} \textbf{24}, 229--236 (1989).

\end{thebibliography}
\end{document}